\documentclass[a4paper,pdftex,floatfix,prl,showpacs,twocolumn,twoside]{revtex4}

\usepackage{dcolumn}
\usepackage{graphicx}
\usepackage{textcomp}
\usepackage{times}
\usepackage{hypernat,hyperref}

\newlength{\figwidth}
\newcommand{\ie}{i.\,e.}
\newcommand{\eg}{e.\,g.}
\newcommand{\mueff}{\ensuremath{\mu_{\text{eff}}}}

\begin{document}

\title{A selector for structural isomers of neutral molecules}%
\author{Frank Filsinger}%
\author{Undine Erlekam}%
\author{Gert von Helden}%
\author{Jochen K\"upper}%
\email[Author to whom correspondence should be addressed. Email:~]{jochen@fhi-berlin.mpg.de}%
\author{Gerard Meijer}%
\affiliation{Fritz-Haber-Institut der Max-Planck-Gesellschaft, Faradayweg 4--6, 14195 Berlin,
   Germany}%
\date{\today}%
\begin{abstract}
   \noindent%
   We have selected and spatially separated the two conformers of 3-aminophenol (C$_6$H$_7$NO)
   present in a molecular beam. Analogous to the separation of ions based on their mass-to-charge
   ratios in a quadrupole mass filter, the neutral conformers are separated based on their different
   mass-to-dipole-moment ratios in an ac electric quadrupole selector. For a given ac frequency, the
   individual conformers experience different focusing forces, resulting in different transmissions
   through the selector. These experiments demonstrate that conformer-selected samples of large
   molecules can be prepared, offering new possibilities for the study of gas-phase biomolecules.
\end{abstract}
\maketitle%

\noindent%
During the last decades, the properties of biomolecules in the gas phase have been studied in ever
greater detail~\cite{EPJD20:Biomolecules, PCCP6:Biomolecules, Vries:ARPC58:585}. Although the study
of biomolecules outside of their natural environment was met with skepticism in the beginning,
spectroscopic studies on isolated species in a molecular beam have proven to be very powerful to
understand their intrinsic properties. Moreover, their native environment can be mimicked by adding
solvent molecules one by one. These studies on well-defined biomolecular systems are particularly
relevant to benchmark theoretical calculations. Even in the cold environment of a molecular beam,
biomolecules exist in various conformational structures. The existence of multiple conformers
(structural isomers) has been observed in the study of jet-cooled glycine for the first
time~\cite{Suenram:JACS102:7180} and in numerous experiments since then. In many cases, the
individual conformers are identified via their different electronic spectra~\cite{Rizzo:JCP83:4819,
   Nir:Nature408:949}. This has been exploited in multiple-resonance techniques to measure, for
instance, conformer-specific infrared spectra from which the conformational structures can be
deduced~\cite{Snoek:CPL321:49, Bakker:PRL91:203003}. Apart from this information on the local minima
on the potential energy surface, information on the barriers separating the conformers has been
obtained in sophisticated multiple-resonance experiments as well~\cite{Dian:Science303:1169}.

The preparation of conformer-selected samples of biomolecules would enable a new class of
experiments to be performed on these systems, \eg, electron and X-ray
diffraction~\cite{Hedberg:Science254:410, Chapman:NatPhys2:839:short} and tomographic imaging
experiments~\cite{Itatani:Nature432:867}. Also, ultrafast dynamics studies on the ground-state
potential energy surface would benefit from the availability of these pure samples. For charged
species the separation of structurally different molecules has been demonstrated using ion-mobility
in drift tubes~\cite{Helden:Science267:1483, Jarrold:PCCP9:1659}. For neutral molecules, no such
separation method exists. It has been demonstrated that the abundance of the conformers in the beam
can be partly influenced by selective over-the-barrier excitation in the early stage of the
expansion~\cite{Dian:Science296:2369} or by changing the carrier gas~\cite{Erlekam:PCCP9:3783}.
These methods, however, are not generally applicable nor able to specifically select each of the
conformers.

Polar molecules experience a force in an inhomogeneous electric field given by the negative gradient
of the Stark energy. If the molecule is in an eigenstate whose Stark energy increases with
increasing electric field, a so-called low-field-seeking (lfs) state, it feels a force towards
regions of low electric field. Molecules in lfs states can be focused using static inhomogeneous
electric fields. This has been used, for example, to create the population inversion that was
essential for the demonstration of the MASER~\cite{Gordon:PR99:1264} and for state-selection of
small molecules for scattering experiments. Furthermore, small molecules in lfs states have been
slowed down and trapped using time-varying electric fields~\cite{Bethlem:PRL83:1558,
   Bethlem:Nature406:491}. Large molecules have a high density of rotational states, and due to the
interaction between these states all of them are high-field seeking (hfs), \ie{}, they feel a force
towards regions of high electric field. Molecules in hfs states can be dynamically focused using the
alternating gradient principle~\cite{Auerbach:JCP45:2160}. This has been demonstrated for ammonia in
hfs states~\cite{Kakati:PLA24:676, Junglen:PRL92:223001}. It has also been applied in the
deceleration of CO~\cite{Bethlem:PRL88:133003, Bethlem:JPB39:R263}, YbF~\cite{Tarbutt:PRL92:173002},
and benzonitrile~\cite{Wohlfart:AG-BN:inprep} molecules in hfs states, as well as in the ac trapping
of ND$_3$~\cite{Veldhoven:PRL94:083001}.

The conformers of a specific biomolecule all have the same mass $m$ and the same connectivities
between the atoms (constitution) but differ by the orientations of their functional groups in the
molecular frame, \ie, by their folding pattern. The vectorial sum of the local dipole moments of the
functional groups largely determines the overall dipole moment of the molecule. The different dipole
moments $\mu$ of the conformers can be exploited to select individual conformers using dynamic
focusing with ac electric fields. This is most easily implemented in a setup using high voltage
electrodes in a quadrupole arrangement around a molecular beam. The operation principle of such an
$m/\mu$-selector is equivalent to that of the $m/q$ quadrupole mass filter for charged particles,
where $q$ is the charge~\footnote{Alternatively, electric Stern-Gerlach-type beam deflection
   experiments~\cite{Broyer:PhysScr76:C135} could, in principle, provide partial spatial separation
   of conformers. However, it would only separate the most polar conformer from the others and would
   not provide an active confinement of the selected conformer. To the best of our knowledge, this
   application of beam-deflection has not been demonstrated yet.}. In the $m/\mu$-selector the most
polar quantum states of a given conformer are focused most efficiently and have the highest
transmission. These are also the quantum states that can be aligned or oriented best, using intense
laser fields or strong static electric fields, respectively. Therefore, experiments that rely on
highly oriented samples, such as tomographic or diffraction imaging experiments, would particularly
benefit from the conformer-selected polar ensembles generated by the $m/\mu$-selector.

In this Letter, we demonstrate the selective transmission of the cis and trans conformers of neutral
3-aminophenol (C$_6$H$_7$NO) through an ac quadrupole $m/\mu$-selector. The cis and trans conformers
of 3-aminophenol are used here as prototypes for the different structural isomers of biomolecules.
The rotational envelopes of the electronic origin transitions of the individual conformers provide
information on the rotational state specific transmission of the device.

\begin{figure}
   \centering
   \includegraphics[width=\figwidth]{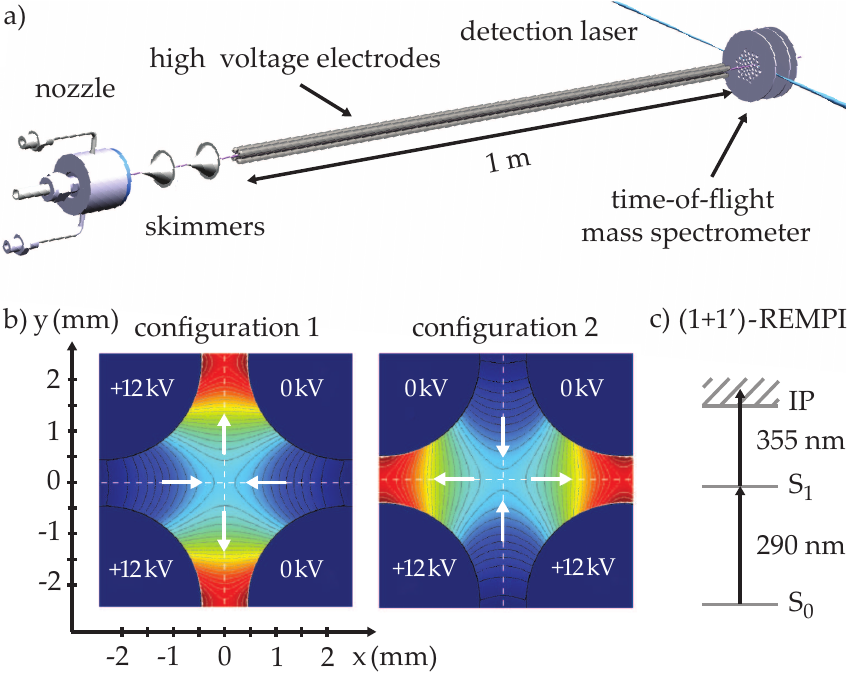}
   \caption{(Color online) a) Scheme of the experimental setup. b) Cut through the high voltage
      electrodes and electric field configurations; red areas represent regions of strong electric
      field, blue areas regions of weak electric field. The electric field strength at the center is
      45~kV/cm and contour lines are given every 4~kV/cm. White arrows indicate the direction of the
      force experienced by molecules in high-field seeking quantum states. c) Two color ionization
      detection scheme.}
   \label{fig:setup}
\end{figure}

The experimental setup is shown in Fig.~\ref{fig:setup}\,a.
A sample of 3-aminophenol is heated to 110~\textdegree{C} and co-expanded in 2~bar of Kr through a
pulsed nozzle operated at a repetition rate of 10~Hz. The mean velocity of the molecules in the beam
is approximately 480~m/s with a velocity spread (full-width at half-maximum) of about 10~\%. After
passing two skimmers, placed 5~cm and 15~cm downstream from the nozzle, the molecules enter a
second, differentially pumped vacuum chamber, in which the $m/\mu$-selector is placed. The selector
consists of four polished, 1~m long cylindrical stainless-steel electrodes of 4~mm diameter. High
voltages of 12~kV against ground are applied as shown in Fig.~\ref{fig:setup}\,b. The gaps are
0.9~mm between two adjacent electrodes and 3~mm between two opposing electrodes, resulting in a
field strength of 45~kV/cm on the centerline and a maximum field strength of 135~kV/cm. Using three
high-voltage switches, the field is rapidly switched ($<$ 1 $\mu$s) between the two electric field
configurations shown in Fig.~\ref{fig:setup}\,b. Switching from one configuration to the other
interchanges the role of the $x$ and $y$ axes, \ie, it interchanges the directions of transverse
focusing and defocusing as indicated by the white arrows in Fig.~\ref{fig:setup}\,b. The resulting
dynamic focusing of neutral molecules, which is very similar to the dynamic focusing of charged
particles~\cite{Lee:AccPhys:2004}, has been described in detail elsewhere~\cite{Auerbach:JCP45:2160,
   Bethlem:JPB39:R263}.

The transmitted 3-aminophenol molecules are ionized 1.21~m downstream from the nozzle using
two-color resonance-enhanced multi-photon ionization, (1+1')-REMPI, as shown in
Fig.~\ref{fig:setup}\,c. Prior to entering the detection region, the molecules have to pass through
a 2~mm diameter aperture positioned on the centerline of the selector. The axis of the
$m/\mu$-selector is tilted against the axis of the incoming molecular beam by 0.3\,\textdegree{},
such that predominantly molecules that are transported through the selector enter the detection
region. The laser beams used for excitation and ionization are unfocused, and have diameters of 4~mm
and 2~mm, respectively. To minimize saturation effects, the energy of the frequency-doubled pulsed
dye-laser for electronic excitation (290~nm) is reduced to 20~$\mu$J/pulse, while the ionization
laser (355~nm) is operated with 5~mJ/pulse. The ions are mass-selectively detected in a
time-of-flight setup. The cis and trans conformers can be selectively detected due to their distinct
$S_1\!\leftarrow\!S_0$ excitation wavenumbers of 34109~cm$^{-1}$ and 34467~cm$^{-1}$, respectively.

\begin{figure}
   \centering
   \includegraphics[width=\figwidth]{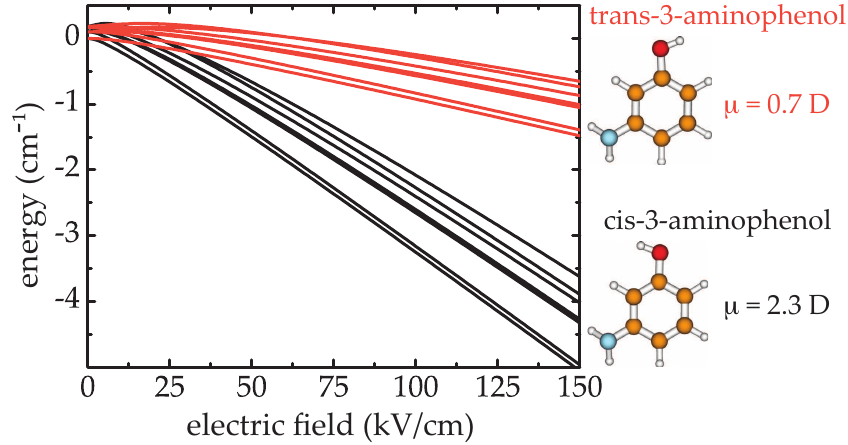}
   \caption{(Color online) Molecular structures, dipole moments, and energies of the lowest
      rotational states of cis- and trans-3-aminophenol as a function of the electric field
      strength.}
   \label{fig:starkshift}
\end{figure}

From the precisely known rotational constants and dipole moments~\cite{Filsinger:PCCP10:666} the
energies of the rotational states of cis-3-aminophenol and trans-3-aminophenol are calculated as a
function of electric field strength. Fig.~\ref{fig:starkshift} shows the resulting Stark curves for
the lowest rotational states of both species.
The transmission characteristics of the selector depend on the effective dipole moment
$\mu_{\text{eff}}$ (the negative of the slope of the Stark curve), the electric field gradients, and
the ac frequency. Similar to the frequency dependence in quadrupole mass-filters, molecules with a
given value of $\mu_{\text{eff}}$ are only transmitted through the selector within a finite range of
frequencies. At too low frequencies molecules are deflected and lost in one transverse dimension
before they are refocused. For high frequencies the time-averaged potential becomes flat resulting
in a strongly reduced transmission. The ac frequency for optimum transmission increases with
increasing \mueff{}. When a constant ac frequency is applied, a \mueff{} selection is performed.

\begin{figure}
   \centering
   \includegraphics[width=\figwidth]{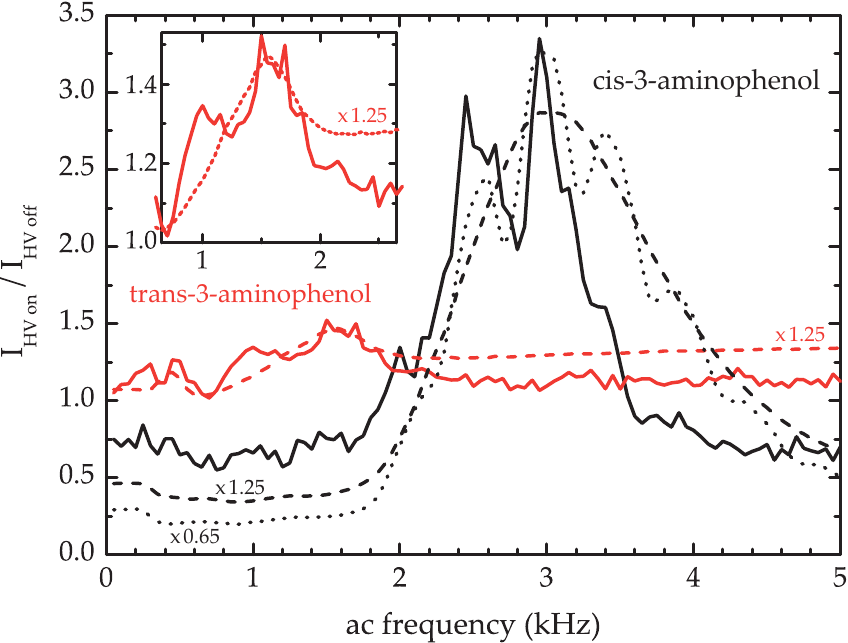}
   \caption{(Color online) Experimental (solid lines) and simulated (dashed lines) transmission as a
      function of ac frequency for cis-3-aminophenol (black) and trans-3-aminophenol (red). The
      inset shows an enlarged view of the transmission curve of trans-3-aminophenol. The black
      dotted line represents simulations for cis-3-aminophenol that include the details of the
      detection process.}
   \label{fig:frequencyscan}
\end{figure}

AC frequency scans for cis-3-aminophenol and trans-3-aminophenol are shown in
Fig~\ref{fig:frequencyscan}.
The transmission measurements are performed with the excitation laser frequency set close to the
band origin of the respective conformer. The ac frequency is scanned from 0~kHz to 5~kHz in steps of
50~Hz. For a given conformer, the number of transmitted molecules is measured with and without
applied high voltages, and the ratio of these two measurements is plotted in
Fig.~\ref{fig:frequencyscan}. The start-phase of the switching cycle determines the overall
transmission of the selector. In all measurements we start with a half-period of focusing along the
horizontal axis (configuration 1). As a consequence, the phase of the switching cycle at the moment
that the molecules exit the selector changes with the applied frequency. For cis-3-aminophenol a
clear enhancement of the transmission is observed for ac frequencies in the range from 2--3.5~kHz,
whereas for trans-3-aminophenol a weaker transmission maximum is observed around 1.5~kHz. Both, the
higher frequency and the higher transmission for cis-3-aminophenol reflect its considerably larger
dipole moment compared to that of the trans conformer. The central dip in the transmission curve at
2.7~kHz is due to effects of the exact phase of the ac switching cycle at the exit of the selector.
This phase determines the shape of the molecular packet in the detection region, and thereby its
overlap with the laser beams.

\begin{figure}
   \centering
   \includegraphics[width=\figwidth]{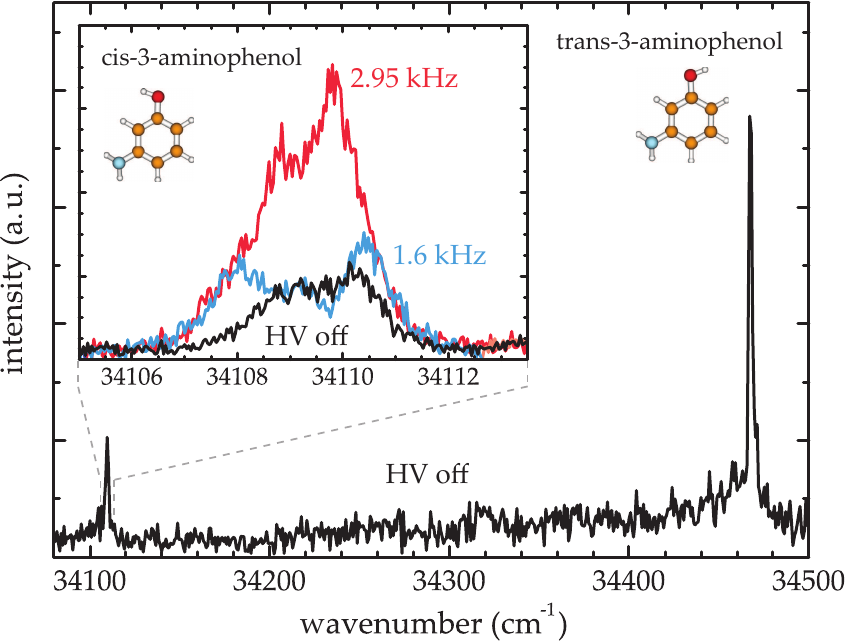}
   \caption{(Color online) (1+1')-REMPI spectrum of 3-aminophenol. The spectrum is measured without
      high voltage applied to the selector, and shows the two origin transitions. In the inset, the
      rotational envelope of the cis-3-aminophenol origin transition is shown without high voltage
      (black) and with high voltage for ac frequencies of 2.95~kHz (red) and 1.6~kHz (blue).}
   \label{fig:spectrum}
\end{figure}

As discussed above, the transmission of the selector depends on the effective dipole moment \mueff{}
of the individual quantum states. Although we cannot detect individual rotational states, there is a
well-defined relation between the excitation laser frequency and the rotational states that are
probed. In Fig~\ref{fig:spectrum} the (1+1')-REMPI spectrum of 3-aminophenol is shown. In the center
of each of the vibronic bands, predominantly transitions from low-$J$ states are probed, whereas the
wings of the rotational envelopes contain mostly transitions from high-$J$ states. The inset of
Fig~\ref{fig:spectrum} shows the rotational contour of the origin transition of cis-3-aminophenol on
an enlarged wavenumber scale measured with and without electric fields, for different ac
frequencies.
For a frequency of 1.6~kHz the wings of the rotational envelope are increased, whereas the intensity
of the central part of the band is actually decreased. This directly reflects that for this ac
frequency high-$J$ states are efficiently transported through the selector, whereas low-$J$ states,
that generally have a larger \mueff{}, are over-focused and have a lower transmission. For a
frequency of 2.95~kHz the whole rotational envelope is clearly increased. The largest enhancement is
now observed for the central part of the rotational envelope, where mostly low-$J$ states are
probed. For trans-3-aminophenol similar, albeit weaker, changes in the rotational envelope are
observed.

Monte-Carlo trajectory calculations are performed to simulate the transmission curves for fixed
excitation laser frequencies. For this, the rotationally resolved electronic excitation spectrum is
calculated using the known rotational constants and transition moment
orientations~\cite{Filsinger:PCCP10:666, Reese:JACS126:11387}. A rotational temperature of 4~K
yields a rotational envelope that agrees best with the observations. A rectangular spectral profile
of the laser with a width of 0.15~cm$^{-1}$ is assumed. For all rotational states that are probed
within this bandwidth of the laser, Monte-Carlo simulations are performed, and individual
transmission curves are calculated. From the calculated line-strengths and populations, a weight for
every single quantum state is determined. The weighted sum over the individual transmission curves
is shown together with the experimental data in Fig.~\ref{fig:frequencyscan} (dashed lines). These
simulations nicely reproduce the peak position and the low-frequency cutoff of the experimental
transmission curves. On the high-frequency side the experimentally observed transmission decreases
faster than predicted, which we attribute to mechanical misalignment. Taking into account the
phase-dependent shape of the molecular packet and its spatial overlap with the detection laser
beams, the observed modulation of the transmission peak for cis-3-aminophenol is correctly
reproduced (dotted line).

In the experiments presented here, the selector is operated under conditions for optimum
transmission, equivalent to the ``rf-only'' operation mode of quadrupole mass filters. The
resolution $\mueff/\Delta\mueff$ of the selector is only about two in this case. In $m/q$ filters,
the resolution $m/\Delta{}m$ is increased by adding a DC offset to the rf potentials, at the cost of
a reduced transmission. In the $m/\mu$-selector a better resolution can be achieved by adding a
static defocusing field to the two configurations of the electric field that we use here. This can
be achieved by using different high voltages for the two electric field configurations or, more
easily, by changing the duty cycle, \ie, by applying the presently used field configurations for
different time-intervals. For many biomolecules, \eg, amino acids and peptides, the various
conformers have large and widely different dipole moments. For phenylalanine, for instance, at least
six conformers have been observed~\cite{Snoek:CPL321:49, Bakker:PRL91:203003} and their dipole
moments are calculated to range from 1~D to 5.5~D. Therefore, the selection of its conformers would
be feasible even at the present resolution.

In summary, we have selected and separated the cis and trans conformers of 3-aminophenol using
switched electric fields in a quadrupole $m/\mu$-selector. The conformers are separated based on
their distinct frequency dependent transmission characteristics. The dynamic focusing works best for
the states with the largest effective dipole moments, which are the lowest rotational states. Here,
molecular packets with an excess of cis-3-aminophenol in low rotational states are created at an ac
frequency of 2.95~kHz. Such conformer-selected molecular packets offer interesting perspectives for
a variety of experiments. Since the selected states are the most polar ones, these samples are
particularly useful for experiments in which aligned or oriented molecules are desired.

\begin{acknowledgments}
   We acknowledge the excellent technical support at the \emph{Fritz-Haber-Institut}. Financial
   support from the \emph{Deutsche Forschungsgemeinschaft} within the priority program 1116
   ``Interactions in ultracold gases'' is acknowledged.
\end{acknowledgments}

\bibliographystyle{apsrev-nourl}
\bibliography{string,mp}

\end{document}